\documentclass[usenatbib]{mnras}
\usepackage{epsfig}
\usepackage{amsmath,amssymb}
\usepackage{graphicx}
\DeclareGraphicsExtensions{.png,.pdf}
\usepackage{txfonts}
\usepackage[english]{babel}
\usepackage[utf8]{inputenc}
\usepackage[T1]{fontenc}
\usepackage{fontenc}
\usepackage{soul}
\allowdisplaybreaks
\def\los/{{line-of-sight}}
\def\Los/{{Line-of-sight}}
\def\av#1{\left\langle#1\right\rangle}

\newcommand\Mstar{M_{\star}}
\newcommand\Mstari{M_{\rm \star,i}}
\newcommand\Mstarlost{M_{\rm \star,lost}}
\newcommand\mstari{m_{\star,i}}
\newcommand\mDMi{m_{{\rm DM},i}}

\newcommand\fstar{f_{\star}}

\newcommand\fDM{f_{\rm DM}}

\newcommand\xiki{\xi_{k,i}}

\newcommand\xistari{\xi_{\star,i}}
\newcommand\xiDMi{\xi_{{\rm DM},i}}

\newcommand\AC{A_{\rm C}}
\renewcommand\AE{A_{\rm E}}
\newcommand\alphaC{\alpha_{\rm C}}
\newcommand\alphaE{\alpha_{\rm E}}
\newcommand\betaC{\beta_{\rm C}}
\newcommand\betaE{\beta_{\rm E}}

\renewcommand\d{{\rm d}}

\newcommand\E{\mathcal{E}}
\newcommand\Etilde{\tilde{\mathcal{E}}}
\newcommand\Etildezero{{\Etilde}_0}
\newcommand\EzeroC{\E_{\rm 0,C}}
\newcommand\EzeroE{\E_{\rm 0,E}}

\newcommand\Ezero{{\E_0}}

\newcommand\Epeak{{\E_{\rm peak}}}

\newcommand\Eq{equation~}

\newcommand\ftot{f_{\rm tot}}
\newcommand\fk{f_k}

\newcommand\Fig{Fig.~}

\newcommand\gammastar{{\gamma_\star}}

\newcommand\Gyr{\,{\rm Gyr}}

\newcommand\Iv{{\boldsymbol I}}
\newcommand\Jv{{\boldsymbol J}}

\newcommand\Lz{L_z}

\newcommand\kms{{\rm \,km\,s^{-1}}}
\newcommand\kpc{{\rm \,kpc}}

\newcommand\M{M}

\newcommand\Mtot{{\M}_{\rm tot}}
\newcommand\Mthreekpc{{\M}_{\rm 3\,kpc}}
\newcommand\Mthreekpci{{\M}_{\rm 3\,kpc,i}}

\newcommand\Msun{{\rm \,M}_{\odot}}

\newcommand\Ncomp{{N}_{\rm comp}}
\newcommand\Nstar{N_{\star}}
\newcommand\Nbin{N_{\rm bin}}
\newcommand\NDM{{N}_{\rm DM}}
\newcommand\Ntot{{N}_{\rm tot}}

\newcommand\pc{{\rm \,pc}}

\newcommand\rhozero{\rho_0}
\newcommand\rhotilde{\tilde{\rho}}
\newcommand\ftilde{\tilde{f}}
\newcommand\rhostar{\rho_{\star}}

\newcommand\rhotot{\rho_{\rm tot}}

\newcommand\rhoDM{\rho_{\rm DM}}

\newcommand\Psitot{\Psi_{\rm tot}}
\newcommand\Phitot{\Phi_{\rm tot}}
\newcommand\PhiDM{\Phi_{\rm DM}}

\newcommand\Phistar{\Phi_\star}

\renewcommand\P{\mathcal{P}}
\newcommand\Pk{\mathcal{P}_k}
\newcommand\Pstar{\mathcal{P}_\star}
\newcommand\PDM{\mathcal{P}_{\rm DM}}
\newcommand\rhalfstar{r_{\rm half,\star}}

\newcommand\rt{r_{\rm t}}

\newcommand\sigmalos{\sigma_{\rm los}}
\newcommand\sigmalosj{\sigma_{{\rm los},j}}
\newcommand\sigmalosjsquared{\sigma_{{\rm los},j}^2}

\newcommand\Sigmastarj{\Sigma_{\star,j}}
\newcommand\sigmalostail{\sigma_{\rm los,tail}}
\newcommand\Sect{Section~}
\newcommand\Sects{Sections~}

\newcommand\tdyn{t_{\rm dyn}}

\newcommand\vy{v_y}
\newcommand\vyi{v_{y,i}}

\newcommand\vv{{\boldsymbol v}}
\newcommand\xv{{\boldsymbol x}}

\newcommand\yr{{\rm \,yr}}

\begin{document}

\date{Accepted 2021 March 10. Received 2021 March 8; in original form 2020 December 11}

\title[$N$-body models of composite stellar systems]{Effective $N$-body models of composite collisionless stellar systems} 

\author[C.\ Nipoti et al.]{\parbox{\textwidth}{Carlo Nipoti$^{1}$\thanks{E-mail: carlo.nipoti@unibo.it}, Giacomo Cherchi$^{1}$, Giuliano Iorio$^{2,3,4}$ and Francesco Calura$^{5}$}\vspace{0.4cm}\\
  \parbox{\textwidth}{
$^1$Department of Physics and Astronomy, University of Bologna,
    via Gobetti 93/2, I-40129 Bologna, Italy \\
    $^2$Physics and Astronomy Department Galileo Galilei, University of Padova, Vicolo dell'Osservatorio 3, I-35122, Padova, Italy\\
$^{3}$INFN-Padova, Via Marzolo 8, I--35131 Padova, Italy\\ 
	$^{4}$INAF-Padova, Vicolo dell'Osservatorio 5, I--35122 Padova, Italy\\      $^5$INAF - Astrophysics and Space Science Observatory of Bologna, via Gobetti 93/3, I-40129, Bologna, Italy \\
}}

\maketitle 

\begin{abstract}
Gas-poor galaxies can be modelled as composite collisionless stellar
systems, with a dark matter halo and one or more stellar components,
representing different stellar populations.  The dynamical evolution
of such composite systems is often studied with numerical $N$-body
simulations, whose initial conditions typically require realizations
with particles of stationary galaxy models. We present a novel method
to conceive these $N$-body realizations, which allows one to exploit
at best a collisionless $N$-body simulation that follows their
evolution. The method is based on the use of an effective $N$-body
model of a composite system, which is in fact realized as a
one-component system of particles that is interpreted {\em a
  posteriori} as a multi-component system, by assigning in
post-processing fractions of each particle's mass to different
components.  Examples of astrophysical applications are $N$-body
simulations that aim to reproduce the observed properties of
interacting galaxies, satellite galaxies and stellar streams.  As a
case study we apply our method to an $N$-body simulation of tidal
stripping of a two-component (dark matter and stars) satellite dwarf
galaxy orbiting in the gravitational potential of the Milky Way.
\end{abstract}

\begin{keywords}
  dark matter -- galaxies: evolution -- galaxies: interactions -- galaxies:
  kinematics and dynamics -- methods: numerical
\end{keywords}

\section{Introduction}
\label{sec:intro}

Collisionless $N$-body simulations are standard tools to study the
evolution of stellar systems such as galaxies and clusters of
galaxies, with typical applications ranging from stability analysis,
to the study of galaxy interactions and mergers, tidal stripping of
satellites, and dynamical friction.  When the phenomenon studied with
the $N$-body simulation involves composite collisionless stellar
systems, to set up the initial conditions it is often necessary to
build $N$-body realizations of stationary multi-component models.
Here we present effective $N$-body models that allow one to study
efficiently the evolution of such composite systems.

The approach presented here can be used in several studies
of galactic dynamics, provided the studied galaxies are gas-poor, so
that they can be modelled as multi-component stellar systems, with
dark matter (DM) halos and one or more stellar components,
representing different stellar populations.  Examples of potential
applications are $N$-body simulations of tidal stripping aimed at
reproducing the observed properties of satellite dwarf galaxies
\citep[e.g.][]{Bat15,Ura15,San18,Ior19} or those of tidal streams
\citep[e.g.][]{Lok10,Die17,Lap18,Vas20} in the Milky Way. But, more
generally, the effective $N$-body models presented here can be used in
$N$-body simulations of dissipationless galaxy mergers
\citep[e.g.][]{Nip03a,Boy06,Fri17} or of the dynamical evolution of
galaxies in clusters of galaxies \citep[e.g.][]{Nip03b,Lap13}.

The method proposed in this paper builds on and bears resemblance with
other techniques previously proposed in the literature. The key of the
effective $N$-body models considered here is to design composite
stellar system starting from the total distribution function (DF) and
then obtain its component by subtraction.  In the literature, there
are a few other studies in which composite stellar systems are built
starting from the total DF or mass density
distribution. \citet{Eva93,Eva94} built axisymmetric composite stellar
systems with total logarithmic or power-law gravitational potential,
starting from the analytic DF of the total distribution. Other authors
\citep{Hio94,Cio09,Cio18,Cio19} used instead the total mass density
distribution as starting point to build multi-component anisotropic
spherical stellar systems.  \citet{Whi80} and \citet{Cio95} used a
technique similar to the one used in this work to build equilibrium
models of isotropic or radially anisotropic spherical stellar systems
with metallicity gradients (see also \citealt{Nip03b} and
\citealt{Nip20}). Within this framework, here for the first time we
exploit the idea of building different components by subtraction from
the total DF to envisage a very effective and general method for
$N$-body modelling. This method allows us to use $N$-body simulations
involving only one-component systems to model the dynamical evolution
of entire families of composite stellar systems, with stars and DM.

The paper is organized as follows. In \Sect\ref{sec:stationary} we
review the properties of stationary composite collisionless stellar
systems and introduce the concept of their effective $N$-body
modelling. In \Sect\ref{sec:dynevo} we extend our view to the
dynamical evolution of such systems, when they are not isolated.
\Sect\ref{sec:isotwocomp} treats in more detail the case of
two-component spherical isotropic systems. In \Sect\ref{sec:app} we
present the application of our method to an $N$-body simulation of
tidal stripping. \Sect\ref{sec:concl} concludes.

\section{Stationary composite collisionless stellar systems}
\label{sec:stationary} 

\subsection{Distribution functions and portion functions}
\label{sec:stat_df} 

Let us consider a stationary composite stellar system with $\Ncomp$
components, in which the $k$-th component has DF $f_k$. The total DF is
$\ftot=\sum_{k=1}^{\Ncomp} f_k$. The total gravitational potential
$\Phitot$ generated by these components satisfies the Poisson equation
\begin{equation}
\nabla^2\Phitot(\xv)=4\pi G\rhotot(\xv),
\label{eq:poisson}
\end{equation}  
where
\begin{equation}
  \rhotot(\xv)=\int \ftot \d^3\vv
  \end{equation}  
is the total mass density distribution, and $\xv$ and $\vv$ are,
respectively, the position and velocity vectors.  We know from Jeans'
theorem \citep[e.g.][]{Bin08} that the DFs of stationary collisionless
stellar systems depend on the phase-space coordinates $(\xv,\vv)$
through $n\leq 3$ integrals of motion $\Iv=I_1,...,I_n$, which are
functions of $(\xv,\vv)$ that are conserved along the orbits.  If we
extract from $\ftot=\ftot(\Iv)$ an orbit with integrals of motion
$\Iv$, the probability that a particle on such orbit belongs to the
$k$-th component is
\begin{equation}
\Pk(\Iv)=\frac{\fk(\Iv)}{\ftot(\Iv)}.
\label{eq:pk}
\end{equation}
In this paper we will refer to the function $\Pk(\Iv)$ as the {\em
  portion function} of the $k$-th component.

\subsection{$N$-body realizations}
\label{sec:nbody}

\subsubsection{Standard multi-component $N$-body model}
\label{sec:standard_nbody}

The standard approach to build an $N$-body realization of a stationary
multi-component stellar system is to represent the $k$-th component
with $N_k$ particles with phase-space coordinates extracted from the
DF $\fk(\Iv)$, with $\Iv=\Iv(\xv,\vv)$: the $j$-th particle
($j=1,...,N_k$) has mass $m_j$, phase-space coordinates
$(\xv_j,\vv_j)$ and integrals of motion $\Iv_j=\Iv(\xv_j,\vv_j$). The
total number of particles is $\Ntot=\sum_kN_k$.  Jeans' theorem
guarantees that each component of this $N$-body realization is
stationary, because its particles are extracted from a DF depending
only on integrals of motions. In this approach, in the $N$-body
realization we assign to each particle a given ``kind'', for instance
``DM particle'' or ``stellar particle'' if it belongs to,
respectively, the DM halo or the stellar component. In a purely
collisionless $N$-body system the orbits of particles are determined
only by gravity and are thus independent of the particle kind and on
the particle mass. This suggests to explore different $N$-body
realizations in which the particles are not labelled as being of a
given kind or belonging to a given component. In the following we
introduce such an alternative approach.

\subsubsection{Effective multi-component $N$-body model}
\label{sec:stat_eff}

Instead of extracting a set of particles for each component, as in the
standard method described above, we can construct an $N$-body
realization of a stationary composite stellar system by extracting
$\Ntot$ particles from the total DF $\ftot(\Iv)$, obtaining for the
$i$-th particle a set of phase-space coordinates $(\xv_i,\vv_i)$ and
corresponding integrals of motion $\Iv_i=\Iv(\xv_i,\vv_i)$
($i=1,...,\Ntot$).  In this way, we do not assign a given particle to
one of the components, but we can nevertheless interpret our system as
multi-component as follows.  Given that $\P_k(\Iv_i)$ is the
probability that the $i$-th particle belongs to the $k$-th component
(\Eq\ref{eq:pk}), the mass contribution of the $i$-th particle to the
$k$-th component is $\xiki m_i$, where $m_i$ is the mass of the $i$-th
particle and $\xiki\equiv \P_k(\Iv_i)$ is the mass fraction of the
$i$-th particle that belongs to the $k$-th component. For instance in
a two-component system with a stellar component (with DF $\fstar$) and
DM component (with DF $\fDM\equiv \ftot-\fstar$), the $i$-th particle
has stellar mass $\xistari m_i$ and DM mass $\xiDMi m_i$, where
$\xistari=\Pstar(\Iv_i)$ and $\xiDMi=1-\xistari$ are, respectively,
its stellar and DM mass fractions, and
$\Pstar(\Iv)\equiv\fstar(\Iv)/\ftot(\Iv)$ is the portion function
(\Eq\ref{eq:pk}) of the stellar component.  For any choice of
$\P_k(\Iv)$ the $k$-th component is univocally defined.  For instance,
the total mass of the $k$-th component is $M_k=\sum_i\xiki m_i$, and
similarly one can compute the mass density and velocity distributions
of the $k$-th component simply by weighting the contribution of the
$i$-th particle by $\xiki m_i$.  If such an $N$-body system is evolved
in isolation, the properties (e.g. density and velocity distributions)
of all its $\Ncomp$ components are time-independent in the
limit\footnote{Of course this is not true, strictly speaking, for
  finite $\Ntot$ because of discreteness effects.} $\Ntot\to\infty$,
because $\ftot(\Iv)$ is the DF of a stationary system and $\Pk(\Iv)$
is a function of the integrals of motions.  The main advantage of this
method with respect to the standard method
(\Sect\ref{sec:standard_nbody}) is that $\Pk(\Iv)$ must not be
specified {\em a priori}, so each simulation can be interpreted in
infinite different ways by assuming $\Pk(\Iv)$ {\em a posteriori}.  Of
course, the aim of $N$-body simulations is to study systems whose
physical properties evolve in time: in the next section we move to
discuss such a case.

\section{Dynamical evolution of composite collisionless stellar systems}
\label{sec:dynevo}

$N$-body simulations are often used to study the dynamical evolution,
in the presence of an external perturbation, of stellar systems that
are initially close to equilibrium. Examples are simulations of the
evolution of satellite stellar systems orbiting within a host stellar
system (for instance satellite galaxies orbiting within a host galaxy)
or simulations of galaxy mergers. In order to illustrate our approach,
let us focus on the case of satellites and consider, for instance, the
simulation of a satellite dwarf galaxy made of stars and DM orbiting
in a host galaxy.  As often done in this kind of simulations, we
assume that the host galaxy is represented simply as a static
gravitational potential, while the satellite is represented with
particles as a two-component $N$-body system (with a stellar component
and a DM halo) that would be in equilibrium if isolated
\citep[e.g.][]{Bat15}.

\subsection{Standard multi-component $N$-body models}
\label{sec:evo_standard}

In the standard method the satellite is set up as a two-component
stationary stellar system with $\Nstar$ stellar particles extracted
from a DF $\fstar$ and $\NDM$ DM particles extracted from a DF $\fDM$,
both in equilibrium in the total gravitational potential of the
satellite $\Phitot=\Phistar+\PhiDM$. The total density distribution of
the satellite is $\rhotot=\rhostar+\rhoDM$, where $\rhostar$ is the
density of the stellar component and $\rhoDM$ is the density of the DM
component.  At the initial time of the simulation the phase-space
coordinates of the centre of mass of the satellite are assigned so
that the satellite is in orbit in the fixed external gravitational
potential of the host galaxy. Due to the tidal interaction with the
gravitational field of the host galaxy, the satellite evolves
modifying the distributions of its components, for instance producing
tidal tails, and losing stellar and DM particles via tidal
stripping. The relative distribution of the dark and stellar
components of the satellite are fixed in the initial conditions, so
the outcome of the simulation is univocal.  To explore the evolution
of a satellite on the same orbit, with the same total distribution
function $\ftot=\fstar+\fDM$, but with different dark and stellar DFs,
a new $N$-body simulation is necessary in this standard approach.

\subsection{Effective multi-component $N$-body models}
\label{sec:dyn_eff}

When the effective multi-component $N$-body modelling is used, the
satellite is set up as a one-component stellar system with $\Ntot$
particles extracted from a DF $\ftot(\Iv)$, with total density
distribution $\rhotot$.  As in the standard approach
(\Sect\ref{sec:evo_standard}), at the initial time of the simulation
the satellite is put in orbit in the fixed external gravitational
potential of the host galaxy, and the evolution of all the particles
is followed for the time spanned by the simulation.  The simulation is
then interpreted, {\em a posteriori}, by assigning to each particle a
stellar mass and a DM mass, by choosing a stellar portion function
$\Pstar(\Iv)$, where $\Iv$ are the integrals of motion of the particle
{\em when the satellite is set up in equilibrium and isolated}. In
practice, if the $i$-th particle has mass $m_i$, its stellar mass is
$\mstari=\Pstar(\Iv_i)m_i$ and its DM mass is
$\mDMi=m_i-\mstari=[1-\Pstar(\Iv_i)]m_i$, where $\Iv_i$ are the values
of the integrals of motion of the $i$-th particle in the isolated
satellite. For given $\Pstar$, from the simulation we can infer the
evolution of the stellar and DM components of the satellite,
separately, for instance measuring the stellar and DM mass loss due to
tidal stripping.  The same simulation can be reinterpreted in infinite
ways by choosing different $\Pstar$.

\begin{figure*}
  \centerline{\psfig{file=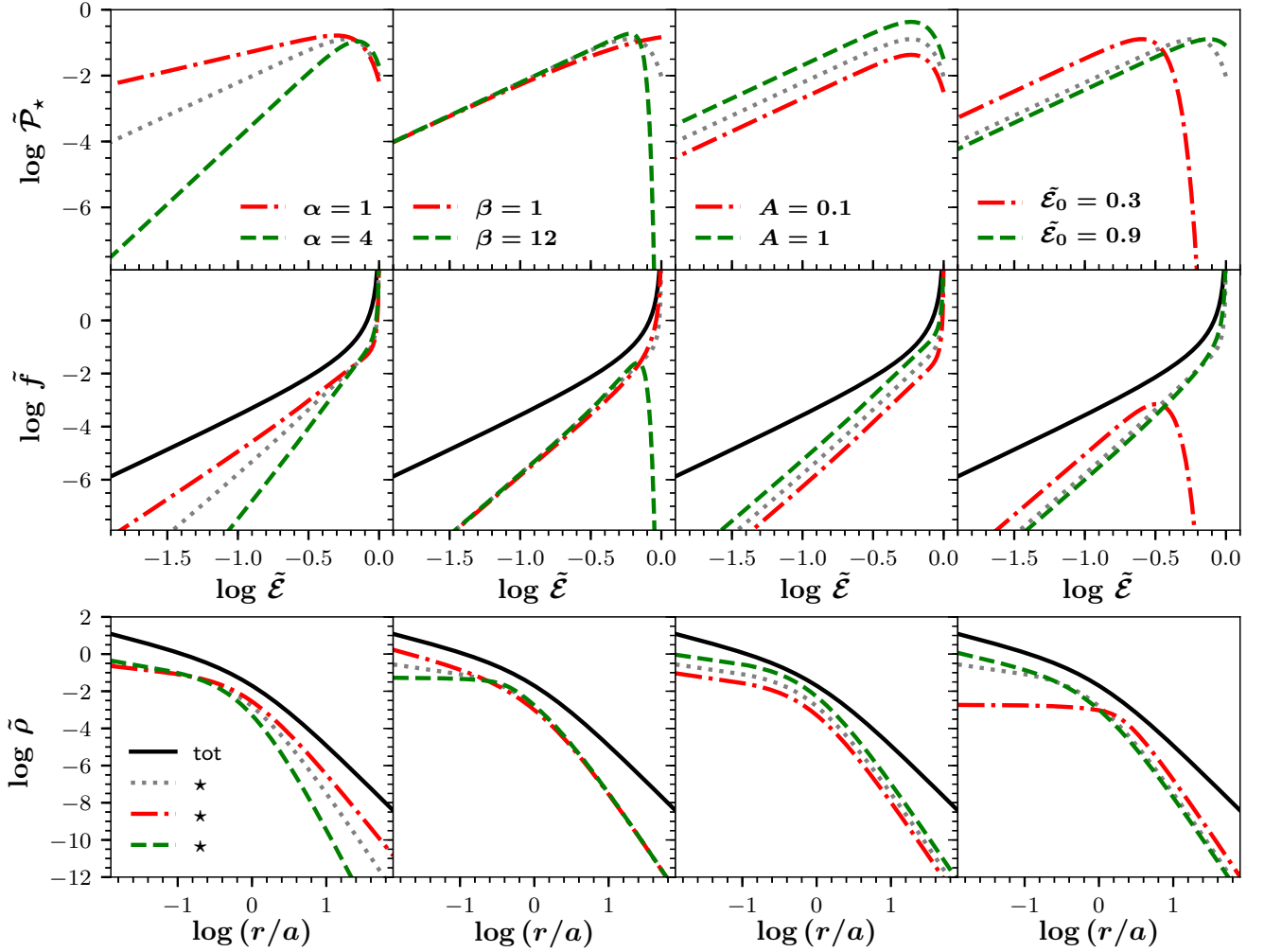,width=\hsize}}
  \caption{Density $\rhotilde\equiv\rho/(\Mtot a^{-3})$ (bottom row of
    panels) as a function of radius and DF $\ftilde \equiv
    f/(G^3M_{\rm tot}a^3)^{-1/2}$ (middle row of panels) as a function
    of specific relative energy $\Etilde\equiv\E/(G\Mtot a^{-1})$ for
    models with the same total distribution (solid curves), but
    stellar distributions (dotted, dashed and dot-dashed curves)
    obtained with different generalized Schechter stellar portion
    functions $\Pstar$ (top row of panels). When not specified
    otherwise, the parameters of the portion function
    (\Eq\ref{eq:por}) are $\alpha=2$, $\beta=4$, $A=0.3$ and
    $\Etildezero\equiv\Ezero/(G\Mtot a^{-1})=0.7$, which are the
    values adopted for the model represented by the dotted curves. In
    each column, the values of the parameters reported in the top
    panel apply also to the middle and bottom panels. $\Mtot$ and $a$
    are, respectively, the total mass and scale radius of the total
    density profile, which is a Hernquist sphere
    (\Eq\ref{eq:den_her}).}
\label{fig:multi_pstar}
\end{figure*}

\section{A simple case: two-component isotropic spherical systems}
\label{sec:isotwocomp}

Here we present an application of the effective $N$-body models
introduced above to spherical two-component collisionless stellar
systems with isotropic velocity distributions.

\subsection{Two-component spherical stellar systems  with ergodic distribution functions}

The simplest family of multi-component collisionless stellar systems
generated by DFs is the family of two-component spherical stellar
systems with isotropic velocity distribution. In this case the DFs of
both components are ergodic, i.e. they are functions only of the
energy per unit mass $E$.  For the sake of clarity, we specialize to
the case in which one of the component is the stellar component, with
DF $\fstar(\E)$, and the other is the DM halo, with DF $\fDM(\E)$,
where $\E=-E$ is the relative energy per unit mass.  The total
distribution function is $\ftot(\E)=\fstar(\E)+\fDM(\E)$.  As
explained in \Sects\ref{sec:stat_eff} and \ref{sec:dyn_eff}, when
building an effective $N$-body model of such a system, we consider a
single component with DF $\ftot(\E)$. The stellar and DM components
are defined by choosing a stellar portion function
$0\leq\Pstar(\E)\leq1$, so $0\leq \fstar(\E)\leq \ftot(\E)$
$\forall\,\E$. The portion function of the DM component is
$\PDM(\E)=1-\Pstar(\E)$, so $0\leq \fDM(\E)\leq \ftot(\E)$
$\forall\,\E$.  One-component systems with the same $\ftot(\E)$ can be
interpreted as different two-component systems, depending on the
choice of $\Pstar(\E)$.  For instance, for an isolated spherical
isotropic system with DF $\ftot(\E)$, the stellar density profile is
\begin{equation}
\rhostar(r)=4\pi\int \Pstar(\E)\ftot(\E) v^2\d v, 
\end{equation}
where $\E(r,v)=\Psitot(r)-\frac{1}{2}v^2$ and $\Psitot(r)=-\Phitot(r)$
is the relative total potential (here $r$ is the spherical radial
coordinate and $v$ the magnitude of the velocity vector). The DM
density distribution is
\begin{equation}
\rhoDM(r)=4\pi\int [1-\Pstar(\E)]\ftot(\E) v^2\d v.
\end{equation}

\subsection{An analytic expression of the portion function}

Our aim is to have an analytic expression of $\Pstar(\E)$, depending
on a handful of parameters, flexible enough to represent realistic
stellar components of spheroids.  In this work we adopt as analytic
expression of the portion function for spherical isotropic systems the
four-parameter function
\begin{equation}
  \Pstar(\E)=A\left(\frac{\E}{\E_0}\right)^\alpha\exp{\left[-\left(\frac{\E}{\E_0}\right)^\beta\right]},
\label{eq:por}  
\end{equation}
where $\alpha$, $\beta$ and $A$ are dimensionless parameters, and
$\E_0$ is a characteristic relative energy.  In the following we will
refer to this analytic function as generalized Schechter function,
because when $\beta=1$ it reduces to the well known \citet{Sch76}
function, widely used in a different context to model the galaxy
luminosity function.  In \Sect\ref{sec:hernq} we show a representative
case in which the generalized Schechter $\Pstar(\E)$ performs well in
producing stellar components with realistic density profiles.
However, we stress that the method proposed in this paper can be
applied with $\Pstar(\E)$ with functional forms different from
\Eq(\ref{eq:por}), for instance with more free parameters if an even
more flexible function is required.

\subsection{A case study: a system with total Hernquist density profile}
\label{sec:hernq}

Let us focus on the case of a self-gravitating system in which the
total density distribution follows a \citet{Her90} profile:
\begin{equation}
  \rhotot(r)=\frac{\Mtot}{2\pi a^3}\frac{1}{(r/a)[1+(r/a)]^3},
\label{eq:den_her}  
\end{equation}
where $a$ is the scale radius and $\Mtot$ the total mass. This total
density distribution is shown in the bottom row of panels of
\Fig\ref{fig:multi_pstar} as a solid curve.  The total gravitational
potential of the system, related to $\rhotot$ by
\Eq(\ref{eq:poisson}), is
\begin{equation}
\Phitot(r)=-\frac{G\Mtot}{r+a}.
\end{equation}
The ergodic DF $\ftot(\E)$ generating a self-gravitating system with
mass density distribution (\ref{eq:den_her}) is know analytically
\citep{Her90} and is shown in the middle row of panels of
\Fig\ref{fig:multi_pstar} as a solid curve.

Such a spherical system with Hernquist total density profile can be
split in a stellar component and a DM component by assuming a stellar
portion function $\Pstar(\E)$.  In particular, adopting as $\Pstar$
the generalized Schechter function (\Eq\ref{eq:por}), we can build
stellar components with double power law density profile, whose
detailed properties depend on the values of the parameters $\alpha$,
$\beta$, $A$ and $\Etildezero\equiv\Ezero/(G\Mtot a^{-1})$.  For
instance, for $\alpha=2$, $\beta=4$, $A=0.3$ and $\Etildezero=0.7$ we
obtain the stellar portion function, DF and mass density distribution
represented by the dotted curves in \Fig\ref{fig:multi_pstar}: the
bottom row of panels shows that the resulting density profile is a
double power law with logarithmic slope $\gammastar\equiv
\d\ln\rhostar/\d\ln r\simeq -0.5$ in the centre and
$\gammastar\simeq-5.5$ in the outskirts. Different slopes can be
obtained by changing the values of the parameters. The parameter
$\alpha$ determines the probability of having weakly bound stars
(i.e.\ with low relative energy $\E$): in particular the lower
$\alpha$ the shallower the outer stellar density profile (see the
leftmost column of panels in \Fig\ref{fig:multi_pstar}). The parameter
$\beta$ determines the probability of having strongly bound stars
(i.e.\ with high $\E$), in the sense that large values of $\beta$
penalize the most bound orbits, thus the higher $\beta$ the shallower
the inner stellar density profile (see the second column of panels in
\Fig\ref{fig:multi_pstar}): in this case a core of constant density is
obtained for $\beta=12$, while for $\beta=1$ $\rhostar\propto r^{-1}$
in the centre.  The parameter $A$, which is the normalization of
$\Pstar$, does not affect the shape of the stellar density profile
but, by shifting vertically $\fstar(\E)$, it determines the fractional
mass contribution of the stellar component, in the sense that the
stars contribute more for higher values of $A$ (see the third column
of panels in \Fig\ref{fig:multi_pstar}).  Finally, the parameter
$\Ezero$ tunes the energy $\Epeak$ at which $\Pstar$ peaks, which for
the generalized Schechter function is
$\Epeak=\Ezero\left(\alpha/\beta\right)^{1/\beta}$. Thus, the value of
$\Ezero$ influences mainly the position of the knee of the stellar
density distribution, which is at larger radii for lower $\Ezero$ (see
the rightmost column of panels in \Fig\ref{fig:multi_pstar}). Note,
however, that also the logarithmic slope $\gammastar$ at radii smaller
than the position of the knee changes with $\Ezero$, because the
stellar DF $\fstar$ (shown in the second row of panels in
\Fig\ref{fig:multi_pstar}) depends not only on $\Pstar$, but also on
the shape of $\ftot$.  The portion function, DF, and density profile
of the DM component, not shown in \Fig\ref{fig:multi_pstar}, can be
obtained simply by subtraction: $\PDM=1-\Pstar$, $\fDM=\ftot-\fstar$
and $\rhoDM=\rhotot-\rhostar$. All these quantities are guaranteed to
be everywhere positive because $\Pstar<1$ $\forall \E$.

\begin{figure}
  \centerline{\psfig{file=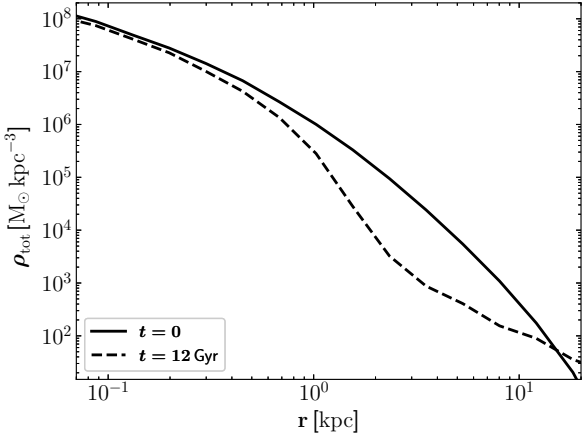,width=\hsize}}
  \caption{Angle-averaged initial ($t=0$, solid curve) and final
    ($t=12\Gyr$, dashed curve) total (DM plus stars) density profiles
    of the satellite in the $N$-body simulation.}
\label{fig:rho}
\end{figure}

\begin{figure*}
  \centerline{
    \includegraphics[trim=40 20 75 10,clip,height=0.3\textwidth]{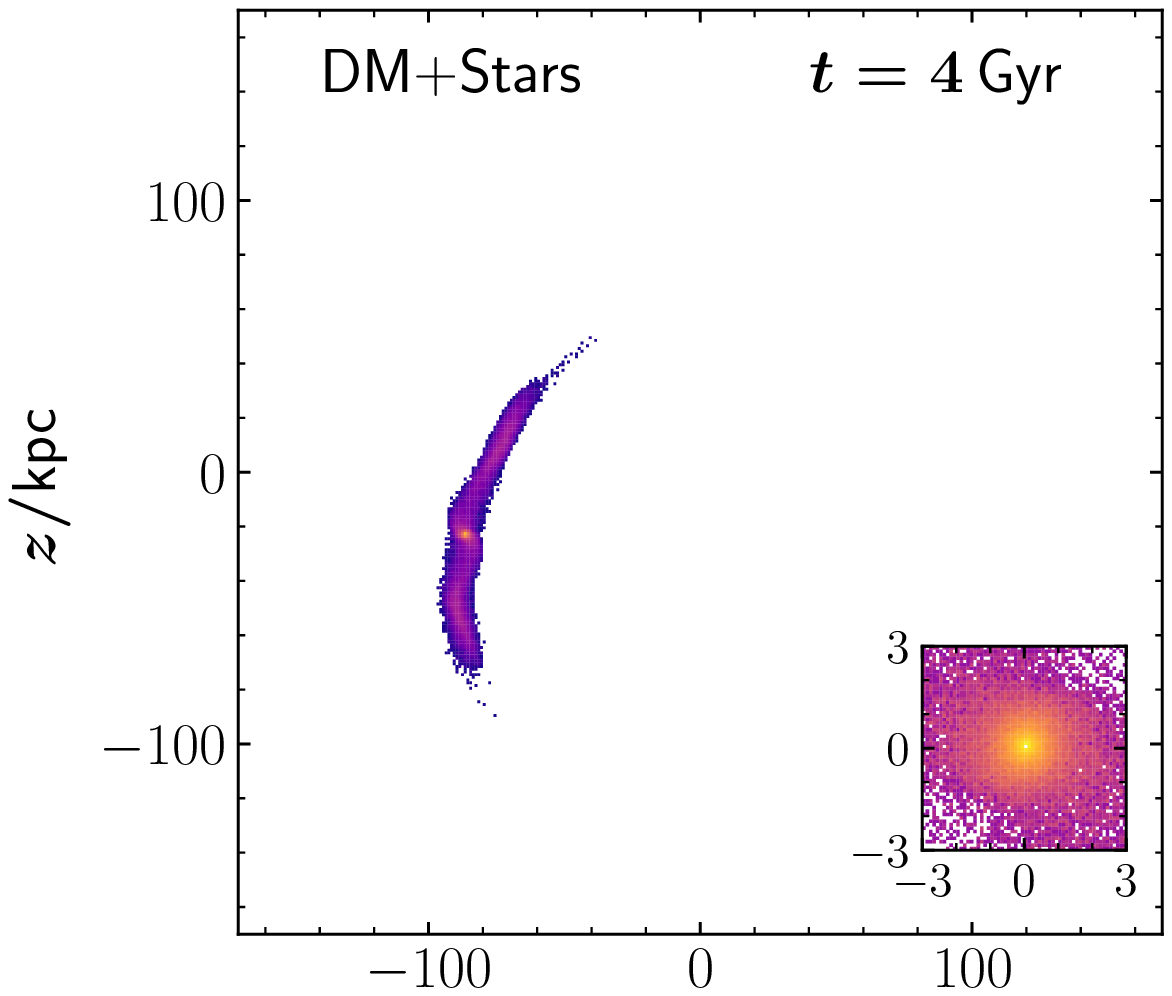}
    \includegraphics[trim=70 20 75 10,clip,height=0.3\textwidth]{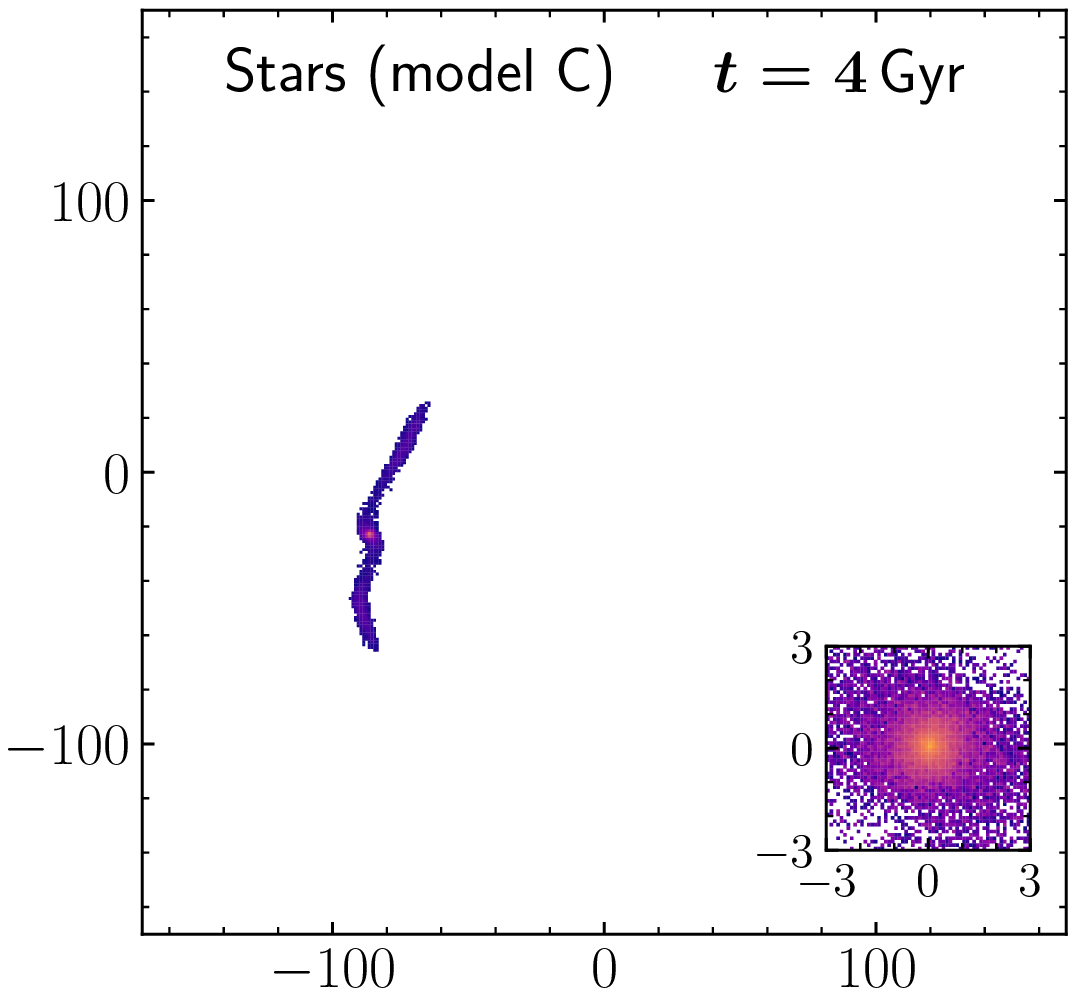}
    \includegraphics[trim=25 20 25 10,clip,height=0.3\textwidth]{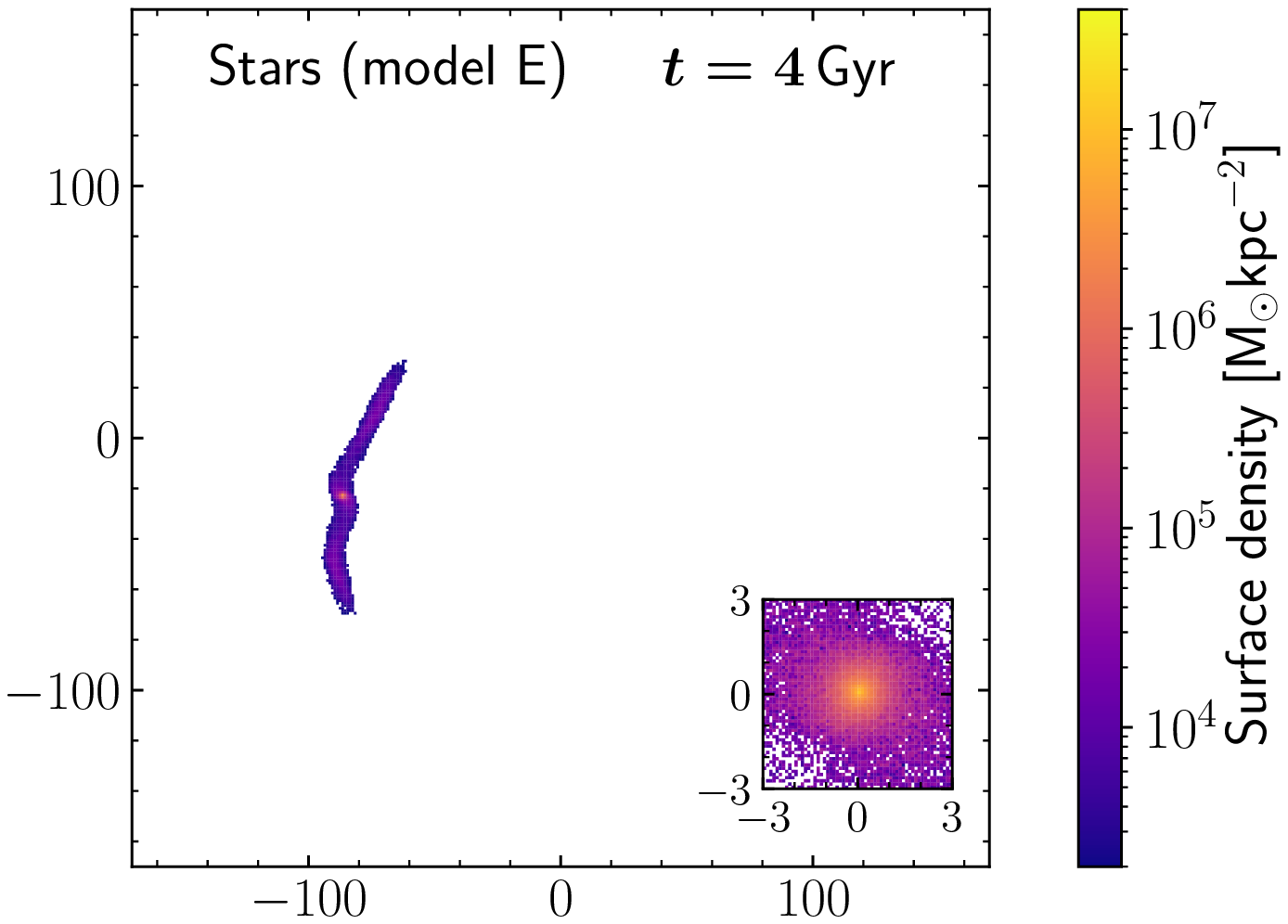}    
  }
  \centerline{
    \includegraphics[trim=40 10 75 20,clip,height=0.3\textwidth]{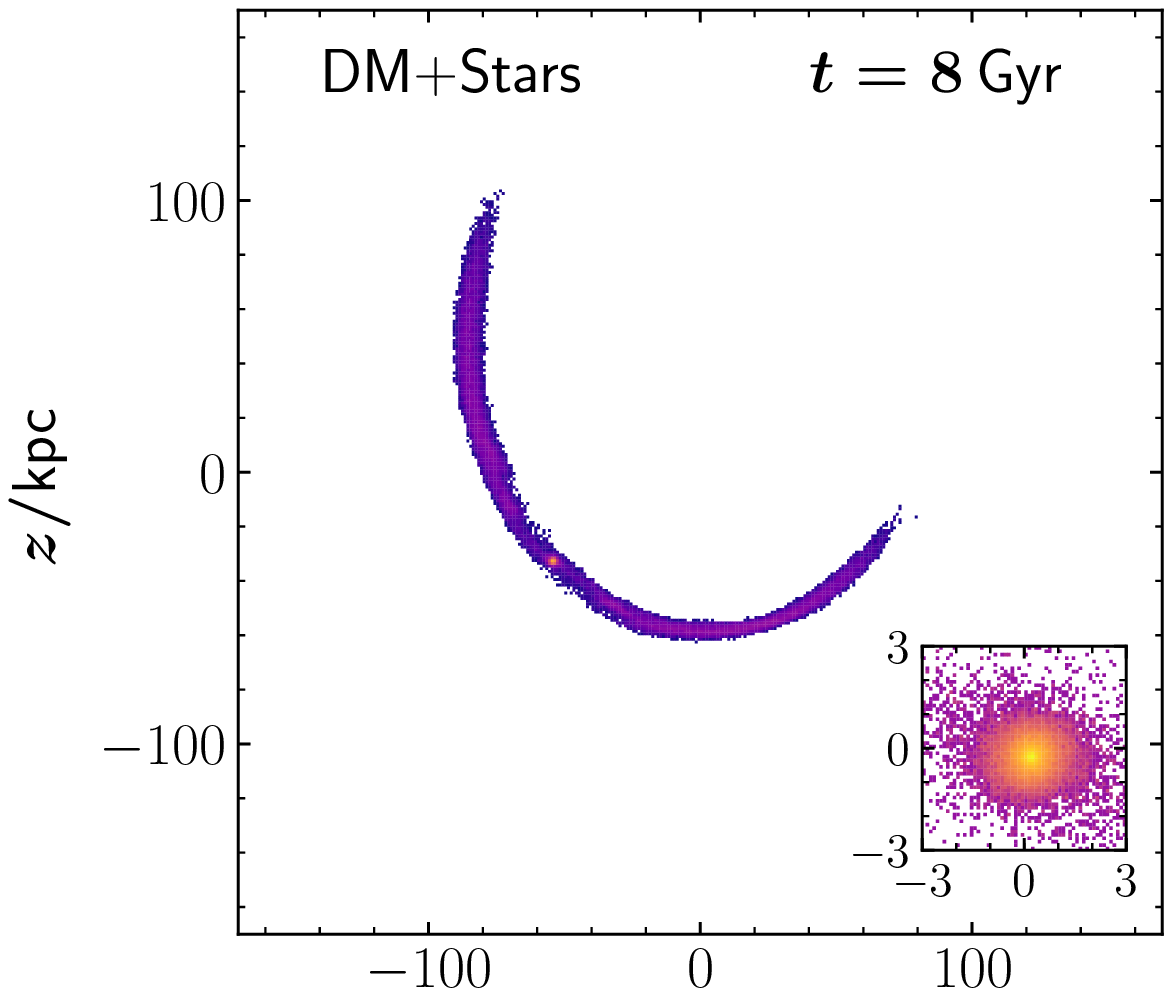}
    \includegraphics[trim=70 10 75 20,clip,height=0.3\textwidth]{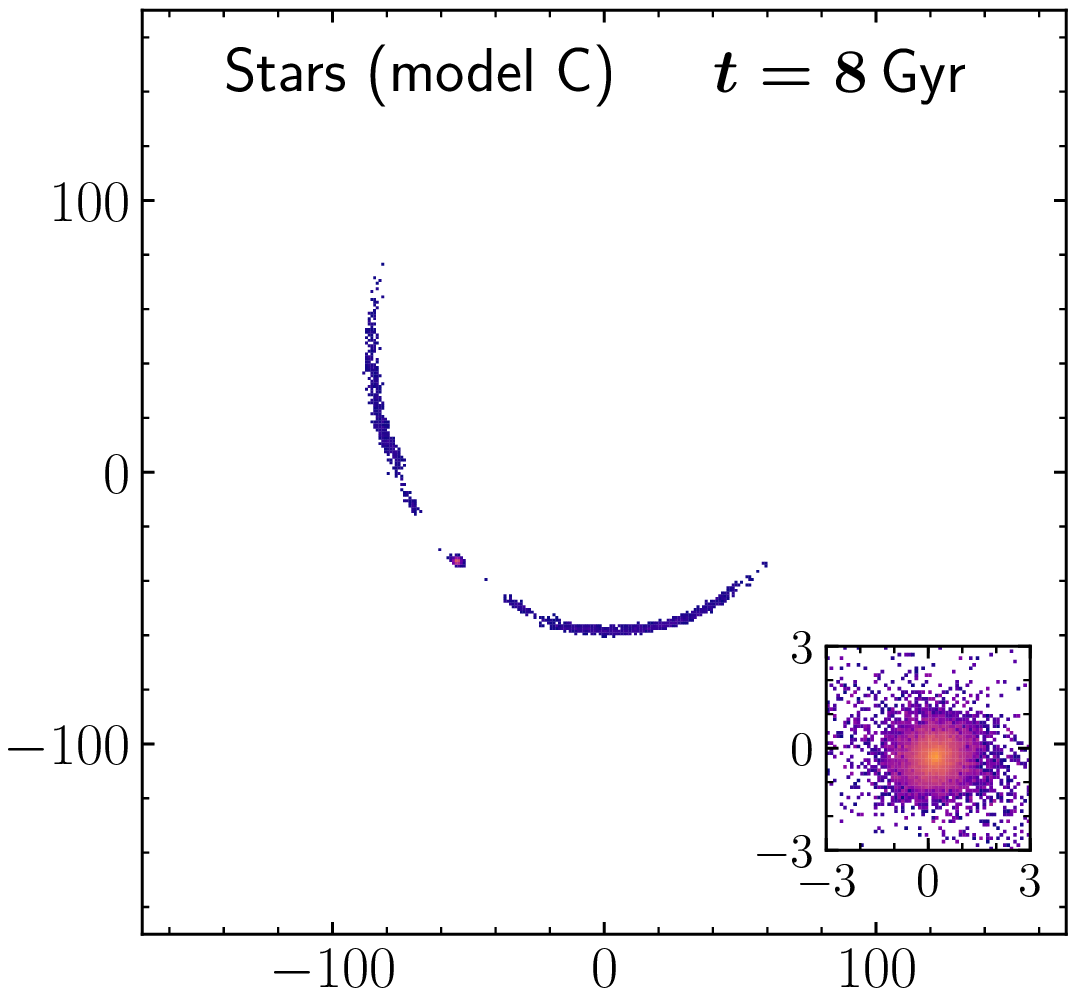}
    \includegraphics[trim=25 10 25 20,clip,height=0.3\textwidth]{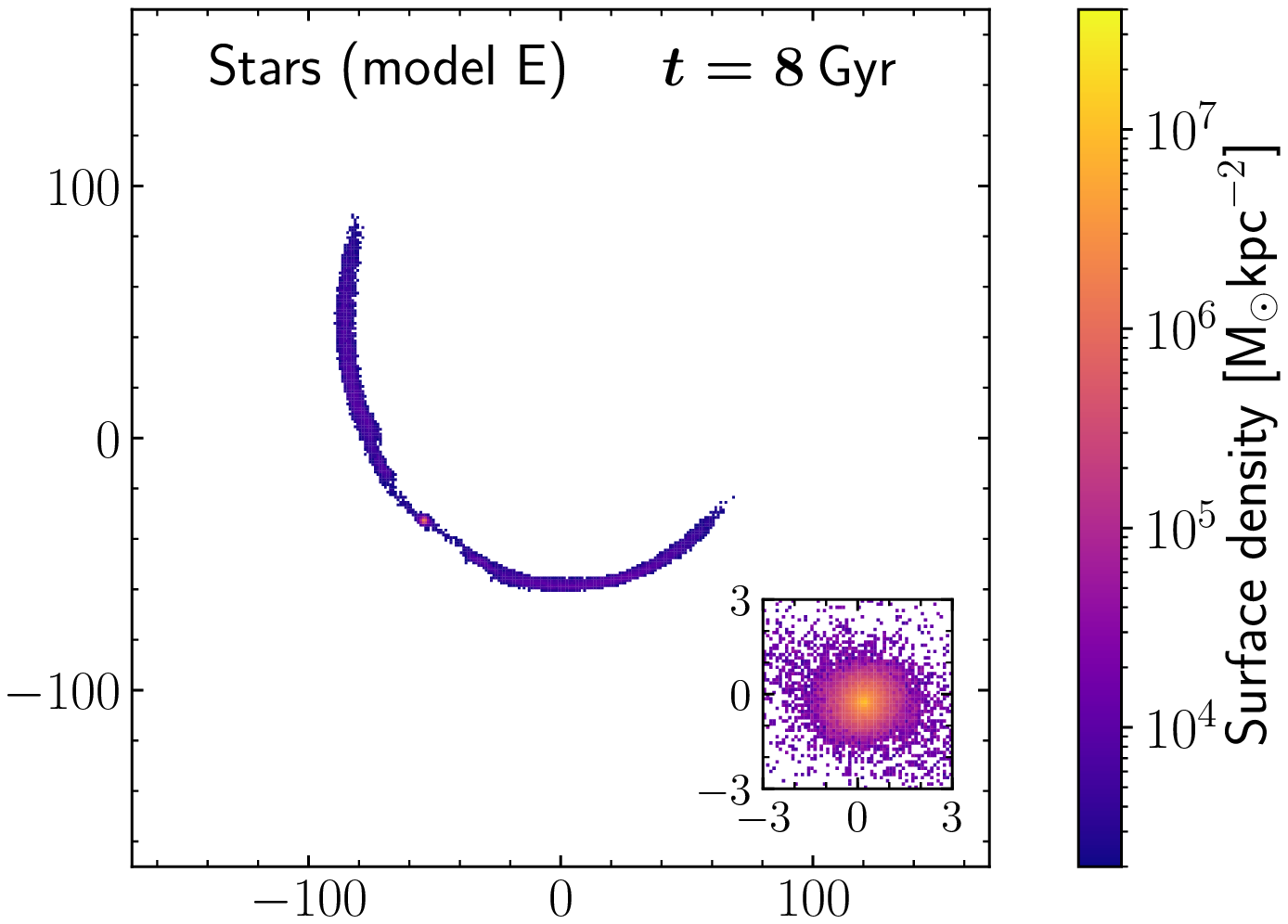}    
  }
  \centerline{
    \includegraphics[trim=40 0 75 30,clip,height=0.3\textwidth]{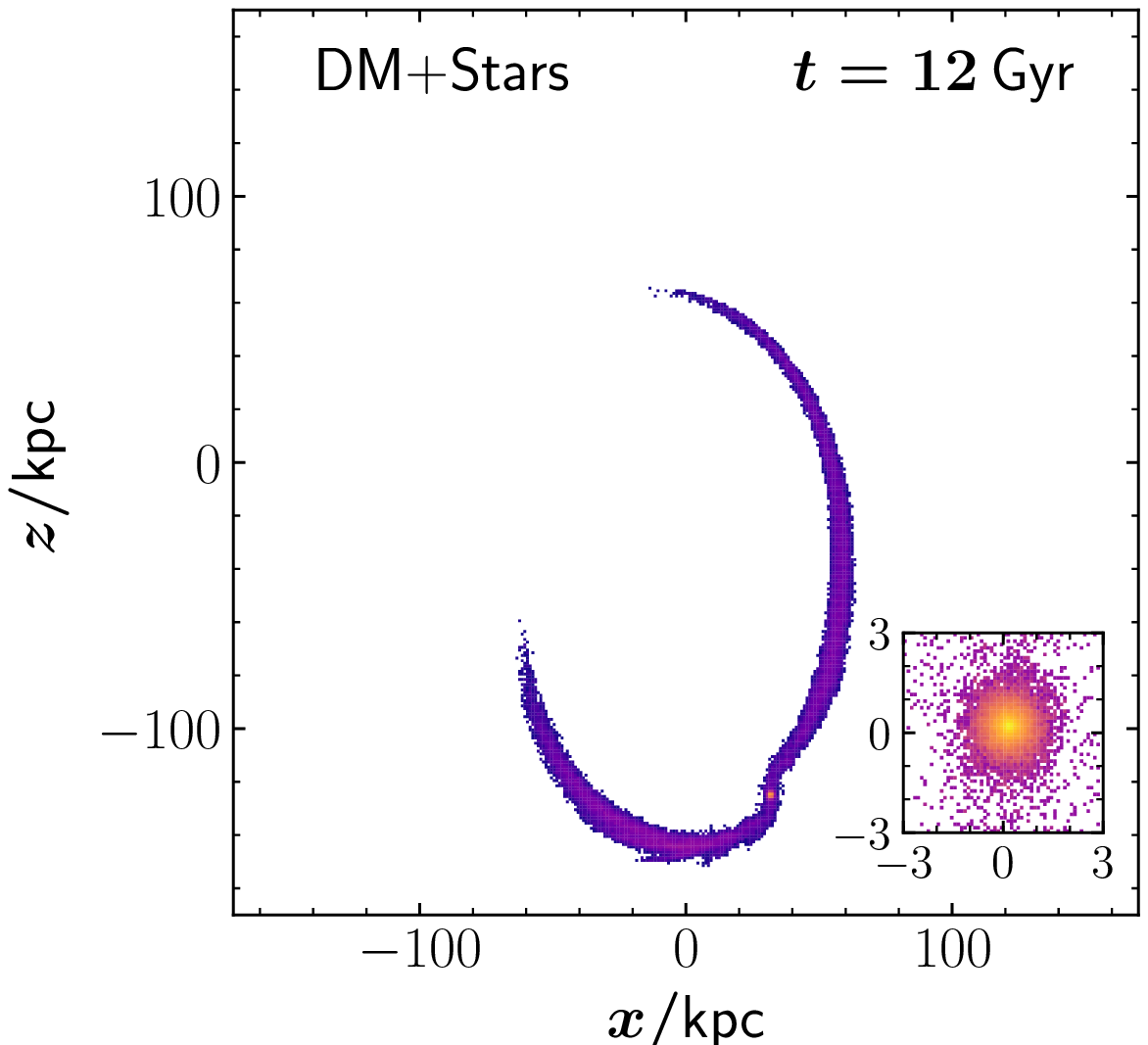}
    \includegraphics[trim=70 0 75 30,clip,height=0.3\textwidth]{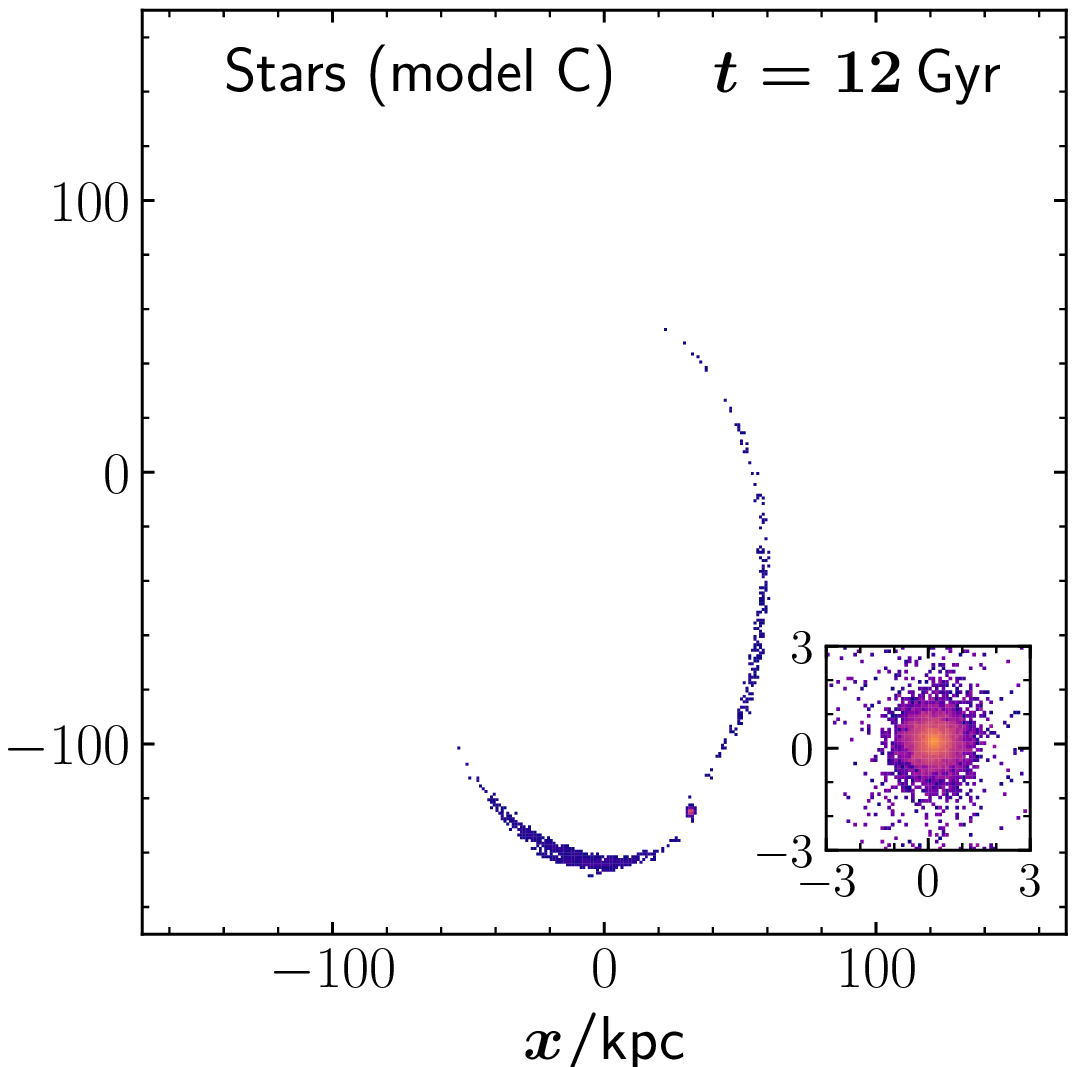}
    \includegraphics[trim=25 0 25 30,clip,height=0.3\textwidth]{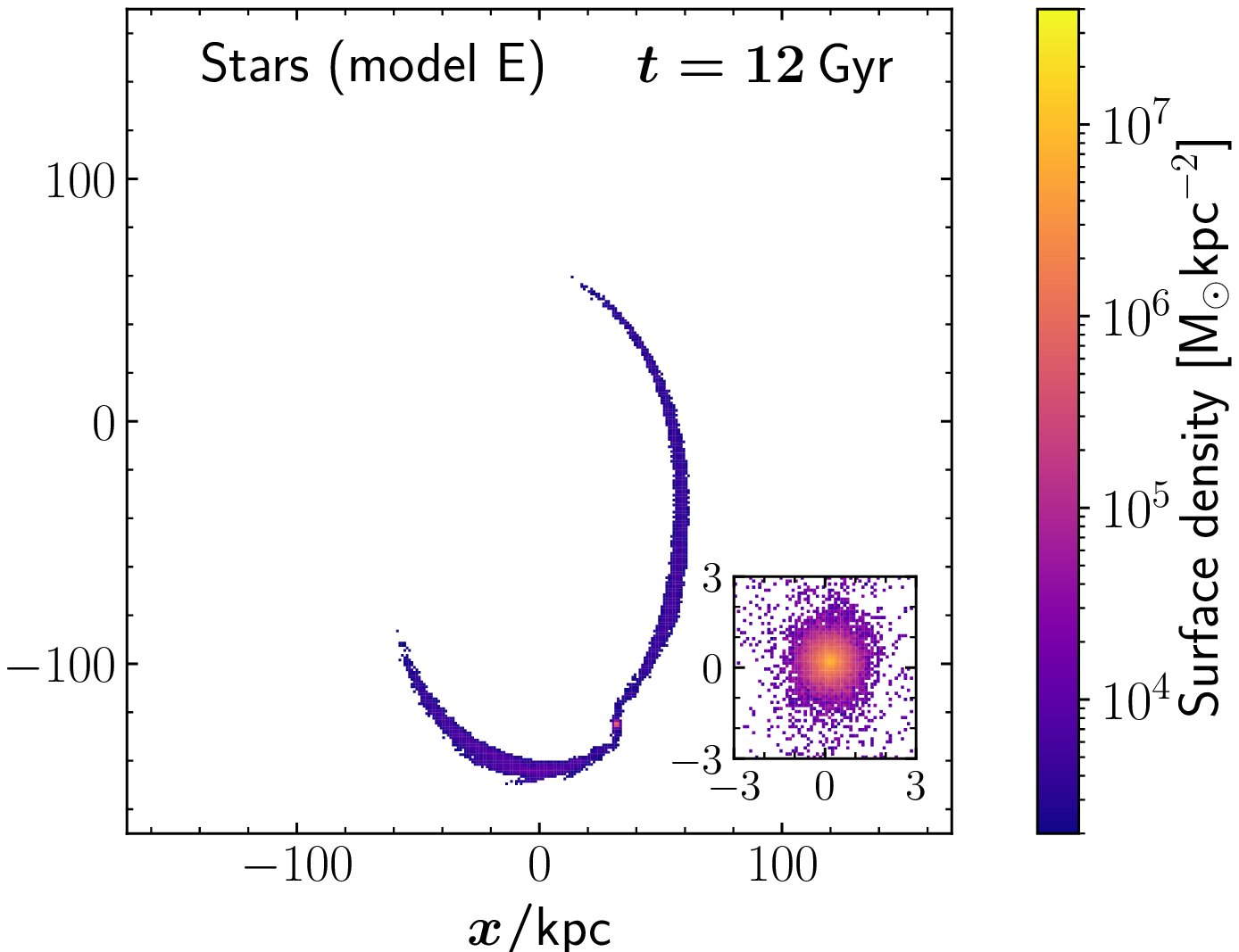}    
  }    
  
 \caption{{\em Left column of panels.} Total (DM plus stars) mass
   surface density distribution of the satellite in the $N$-body
   simulation at $t=4\Gyr$ (top panel), $t=8\Gyr$ (middle panel) and
   $t=12\Gyr$ (bottom panel), for a line of sight along the $y$ axis,
   in the adopted Cartesian coordinate system, centred in the Galactic
   centre, in which the $z$ axis is orthogonal to the Galactic
   equatorial plane $xy$.  {\em Middle column of panels.} Same as left
   column of panels, but showing the stellar mass surface density
   distribution of the satellite according to model C.  {\em Right
     column of panels.} Same as middle column of panels, but for model
   E. In each panel the inset represents a zoomed-in surface density
   map of $3\kpc\times3\kpc$ centred in the peak of the density
   distribution of the satellite.}
\label{fig:snap}
\end{figure*}

\begin{figure}
  \centerline{\psfig{file=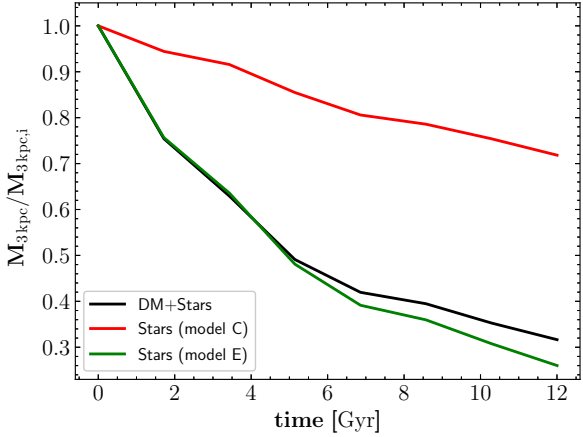,width=\hsize}}
  \caption{Evolution of the total (DM plus stars) mass (black curve)
    and of the stellar mass (red curve for model C and green curve for
    model E) of the satellite in the $N$-body simulation. Here
    $\Mthreekpc$ is the mass within $3\kpc$ from the satellite's
    centre and $\Mthreekpci$ is the initial value of $\Mthreekpc$.}
\label{fig:massloss}
\end{figure}

\begin{figure*}
  \centerline{\psfig{file=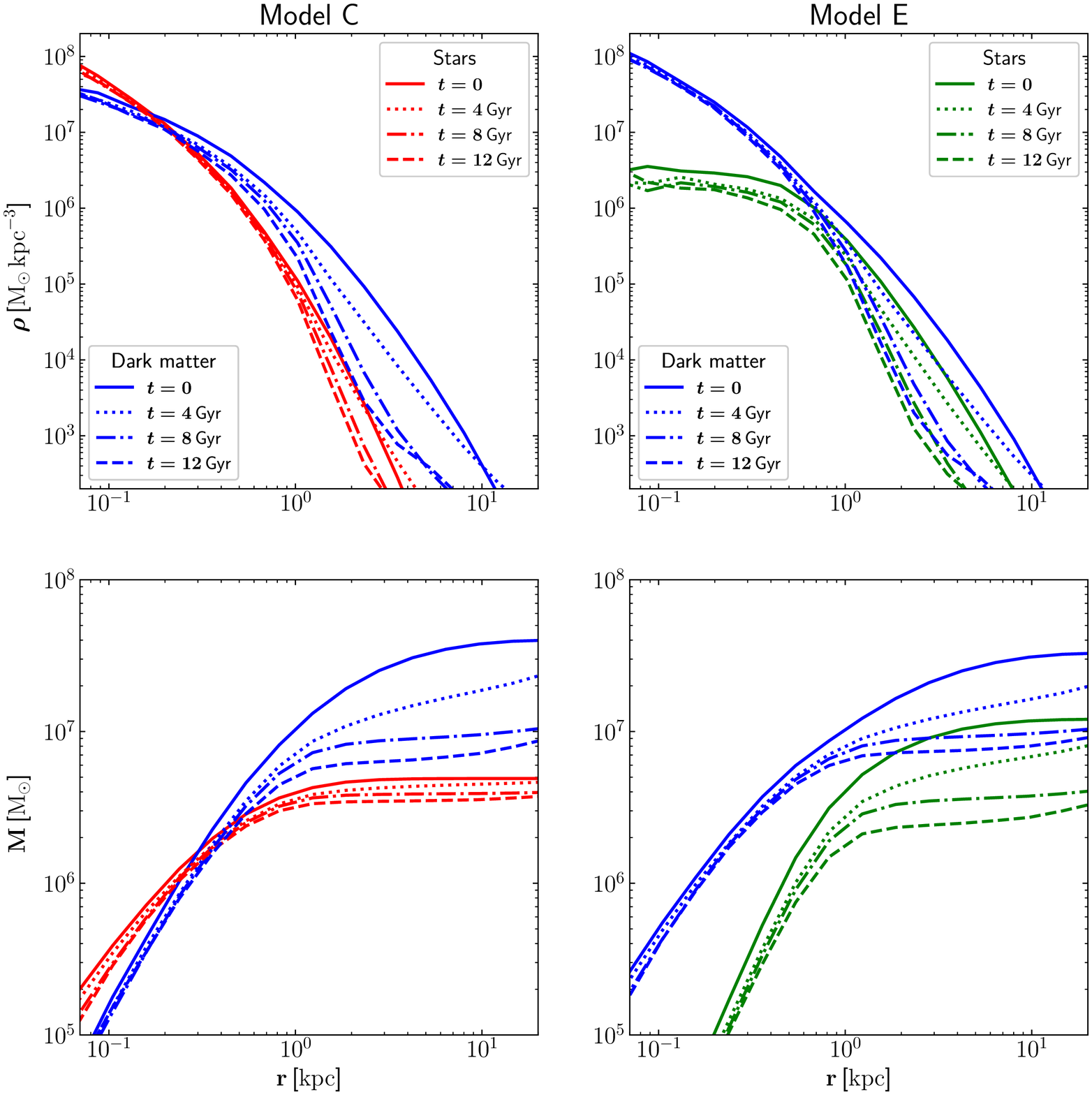,width=\hsize}}
  \caption{{\em Upper panels.} Angle-averaged density profiles of the
    satellite in the $N$-body simulation at different times, indicated
    in the legend, for models C (left panels) and E (right
    panels). The red and green curves represent the stellar density,
    while the blue curves represent the DM density. {\em Lower
      panels.} Stellar and DM mass profiles for the same times and
    models as in the corresponding upper panels.}
\label{fig:evo}
\end{figure*}

\section{Application to an $N$-body simulation of tidal stripping}
\label{sec:app}

Here, we apply the effective multi-component method described above to
an $N$-body simulation that follows the evolution of a satellite
galaxy in the gravitational potential of the Milky Way.

\subsection{Set-up of the $N$-body simulation}
\label{sec:setup}

The initial conditions of the $N$-body realization of the satellite
have been produced using the Python module
\texttt{OpOpGadget}\footnote{https://github.com/iogiul/OpOpGadget}
developed by G.\ Iorio. The $N$-body system is realized as a
one-component spherical isotropic stellar system with density profile
\begin{equation}
   \rhotot(r)=\frac{\rhozero}{(r/a)[1+(r/a)]^3}\exp{\left[-\left(\frac{r}{\rt}\right)^2\right]},
\label{eq:den_her_trunc}   
\end{equation}
representing the total (DM plus stellar) distribution of the
satellite, which is a Hernquist profile (\Eq\ref{eq:den_her})
exponentially truncated at $\rt$. In particular, we adopt $a=0.9\kpc$,
$\rt=17\kpc$ and central density $\rhozero$ such that the total mass
of the system is $\Mtot\equiv4\pi\int_0^\infty \rhotot(r) r^2\d
r=4.5\times10^7\Msun$. The satellite's initial total density
distribution in physical units is shown in \Fig\ref{fig:rho} as a
black solid line.  The number of particles is $\Ntot=10^5$, and all
particles have the same mass $m=\Mtot/\Ntot=450\Msun$. The positions
and velocities of the $\Ntot$ particles are assigned in Cartesian
coordinates (relative to the satellite's centre of mass) as in
\citet{Ior19}, using the ergodic DF $\ftot(\E)$ obtained numerically
via Eddington's inversion formula \citep{Edd16}. The $N$-body system
is in equilibrium if isolated, as we verified by running a simulation
with the same initial conditions as that presented in this work, but
with the satellite in isolation, i.e.\ without the Milky Way external
potential.

The simulation was run using the collisionless code \texttt{FVFPS}
\citep{Lon03,Nip03a} with the addition of the axisymmetric Milky Way
model of \citet{Joh95} as external static gravitational potential
\citep[see][]{Bat15}.  We adopted $\theta_{\rm min}=0.5$ as the
minimum value of the opening parameter, softening length
$\epsilon=0.02\kpc$ and constant time step $\Delta t= 0.01 \tdyn$,
where $\tdyn = 1/\sqrt{G \bar{\rho}_{\rm h}}$ is the initial dynamical
time of the satellite and $\bar{\rho}_{\rm h}$ is its initial average
density within the stellar half-mass radius $r_{\rm h}$. For the
adopted initial conditions $\tdyn\simeq 3.5\times 10^8\yr$.

As orbit of the satellite we assume the orbit dubbed P07ecc in
\citet{Bat15}, which is almost polar with eccentricity $\simeq0.4$ and
pericentric radius $\simeq 61\kpc$. At the initial time of the
simulation the phase-space coordinates of the centre of mass of the
satellite are $(x,y,z)=(35.814,\ 0,\ 137.389) \ \kpc$ and
$(v_x,v_y,v_z)=(-94.875, -77.81, 2.901 ) \ \kms$, in a Cartesian
coordinate system, centred in the Galactic centre, in which $xy$ is
the Galactic equatorial plane.  The simulation is evolved for
$12\Gyr$.  For each snapshot of the simulation we measure the
angle-averaged density distribution $\rhotot(r)$ and integrated total
mass distribution $M(r)$, by binning the particles in concentric
spherical shells. Here $r$ is the distance from the satellite's
centre, which is defined as the position of the peak of the density
distribution of the satellite, computed as in \citet{Ior19}. In a
similar way, for given stellar portion function $\Pstar$, we can
measure for each snapshot the angle-averaged stellar density
distribution $\rhostar(r)$ and stellar mass profile $\Mstar(r)$, by
weighting the particles' masses as described in
\Sect\ref{sec:stat_eff}. The DM density and mass distributions are
obtained using as portion function $\PDM=1-\Pstar$.

\subsection{Results}

\subsubsection{Evolution of the total mass distribution}

The projected total (DM plus stars) density distribution of the
satellite at different times in the simulation is shown in
\Fig\ref{fig:snap} (left column of panels), for a line of sight
parallel to the equatorial plane of the Milky Way. As expected, the
initially spherical density distribution of the satellite is distorted
by the interaction with the tidal force field of the Milky Way, which
produces two significant tidal tails, one leading and one trailing,
departing from the main body of the disrupting satellite. However, as
illustrated by the zoomed-in surface density maps in the insets in
\Fig\ref{fig:snap}, the central regions remain close to spherical
symmetry.  While the central total density profile hardly evolves, at
larger radii the total density profile changes drastically with time,
and at $t=12\Gyr$ (black dashed curve in \Fig\ref{fig:rho}) it is
heavily truncated at $r\approx 1\kpc$ and characterized by a shallow
tail at $r\approx 10\kpc$ produced by the stripped particles.  To
quantify the mass loss we take as reference mass at each time the mass
$\Mthreekpc$ of all the particles within a sphere of radius $r=3\kpc$
from the centre of the satellite. The choice of $3\kpc$ as reference
radius (which is about twice the initial half-mass radius) is somewhat
arbitrary, but is empirically motivated by the requirement to include
most of the stellar mass at $t=0$ (see \Sect\ref{sec:evostellardm})
and to exclude most of the stellar tidal tails in the subsequent
snapshots (see insets in \Fig\ref{fig:snap}).  We note that
$\Mthreekpc\simeq 0.69\Mtot$ at $t=0$. The black curve in
\Fig\ref{fig:massloss}, which plots $\Mthreekpc$ as a function of
time, shows that, within $3\kpc$, the satellite loses almost $70\%$ of
its initial mass over 12 Gyr of evolution.

\subsubsection{Evolution of the stellar and dark matter mass distributions}
\label{sec:evostellardm}

The simulation is interpreted {\em a posteriori} in different ways by
choosing different portion functions $\Pstar(\E)$, where $\E$ is the
{\em initial} particle relative energy, computed for the isolated
satellite. Here we consider two models: {\em model C}, in which the
initial stellar distribution is more {\em compact}, and {\em model E},
in which the initial stellar distribution is more {\em extended}. Both
models are obtained assuming as functional form of $\Pstar$ the
generalized Schechter function (\Eq\ref{eq:por}). The values of the
parameters of $\Pstar$ are $\alpha=3$, $\beta=1$, $A=1.35$, and
$\Etildezero=0.8$ for model C, and $\alpha=0.5$, $\beta=12$, $A=0.5$
and $\Etildezero=0.7$ for model E.  The initial stellar density
profile of model C (red solid curve in upper left panel of
\Fig\ref{fig:evo}) has a central cusp ($\rhostar\propto r^{-1}$) and
declines steeply in the outer parts, while the stellar density profile
of model E (green solid curve in upper right panel of
\Fig\ref{fig:evo}) has a central core ($\rhostar\propto {\rm constant}$)
and is shallower in the outskirts. The position of the knee of the
stellar density profile (i.e.\ the radius of transition between inner
and outer slope) occurs at larger radius for model E than for model C.

%
%

The stellar and DM density and mass profiles at different times in the
simulation are shown in \Fig\ref{fig:evo} for model C in the left
column of panels and for model E in the right column of panels.  In
model E the initial DM density is higher than the initial stellar
density at all radii. In model C the initial stellar density is higher
than the DM density in the centre ($r\lesssim 200\pc$), while the dark
halo dominates at larger radii. In both cases the evolution of the DM
density profile resembles that of the total mass distribution, with
substantial losses at large radii.  The evolution of the stellar
component is instead very different in the two cases: the stellar
distribution of model C remains almost unaltered for 12 Gyr, while it
is heavily stripped in model E. The fractional stellar mass loss for
the two models is quantified in \Fig\ref{fig:massloss} using as
reference the stellar mass within a sphere of radius $3\kpc$ from the
satellite's centre: over 12 Gyr in model C the satellite loses about
$30\%$ of its stellar mass, while in model E more than $70\%$ of the
stellar mass is tidally stripped along the orbit.  We note that the
reference radius $r=3\kpc$ encloses the large majority of the stellar
mass at $t=0$ ($98\%$ in model C and $78\%$ in model E), but is small
enough to exclude most of the tidal tails during the orbital
evolution.

The extent, density and morphology of the stellar tidal tails can be
assessed by looking at \Fig\ref{fig:snap}, showing, for models C
(middle column of panels) and E (right column of panels), the
projected stellar density distribution of the satellite at different
times in the simulation for a line of sight parallel to the Galactic
equatorial plane.  The stellar streams are extremely tenuous in model
C, while are much more pronounced in model E.

\subsubsection{Stellar kinematics}
\label{sec:stellarkin}

{Here we study the stellar kinematics of the satellite and of the
  streams, focusing in particular on the line-of-sight stellar
  velocity dispersion $\sigmalos$. As an illustrative case, we take as
  line of sight the direction of the $y$ axis in our reference
  Galactic Cartesian coordinate system and we take as fiducial
  boundary between the main body of the satellite and the tidal tails
  $R=3\kpc$, where $R=\sqrt{(x-x_0)^2+(z-z_0)^2}$ is the projected
  distance from the satellite's centre $(x_0,z_0)$ in the $xz$
  plane. For the main body of the satellite we compute $\sigmalos$
  from sets of particles belonging to circular annuli: the
  line-of-sight velocity dispersion $\sigmalosj$ of the $j$-th radial
  bin $R_j<R<R_{j+1}$ is given by
\begin{equation}
    \sigmalosjsquared=
    \frac{\sum_i\mstari \left(\vyi-\av{\vy}\right)^2}
         {\sum_i\mstari},
  \end{equation}  
where $\vyi$ is the $y$ component of the velocity of the $i$-th
particle,
$\av{\vy}=\left(\sum_i\mstari\vyi\right)/\left(\sum_i\mstari\right)$,
and the sums are over all particles with $R_j<R_i<R_{j+1}$. Here
$\mstari=\Pstar(\E_i)m_i$, where $m_i$ is the mass of the $i$-th
particle and $\E_i$ is its initial energy in the isolated satellite.
\Fig\ref{fig:sigmalossat} shows the satellite's initial and final
profiles of $\sigmalos$ for models C and E.  In the initial conditions
$\sigmalos$ is higher for model E, which has a flatter stellar density
profile, than for model C, which has a steeper stellar density
profile (see upper panels of \Fig\ref{fig:evo}). This just reflects
the fact that, for a given gravitational potential, a higher velocity
dispersion is needed to maintain in equilibrium a more extended
stellar component.  For both models the final $\sigmalos$ profile has
a shape similar to the corresponding initial profile, but lower
normalization: $\sigmalos$ decreases with time mainly because the
potential well becomes shallower, owing to substantial mass loss (see
\Fig\ref{fig:evo}, lower panels).

\begin{figure}
  \centerline{\psfig{file=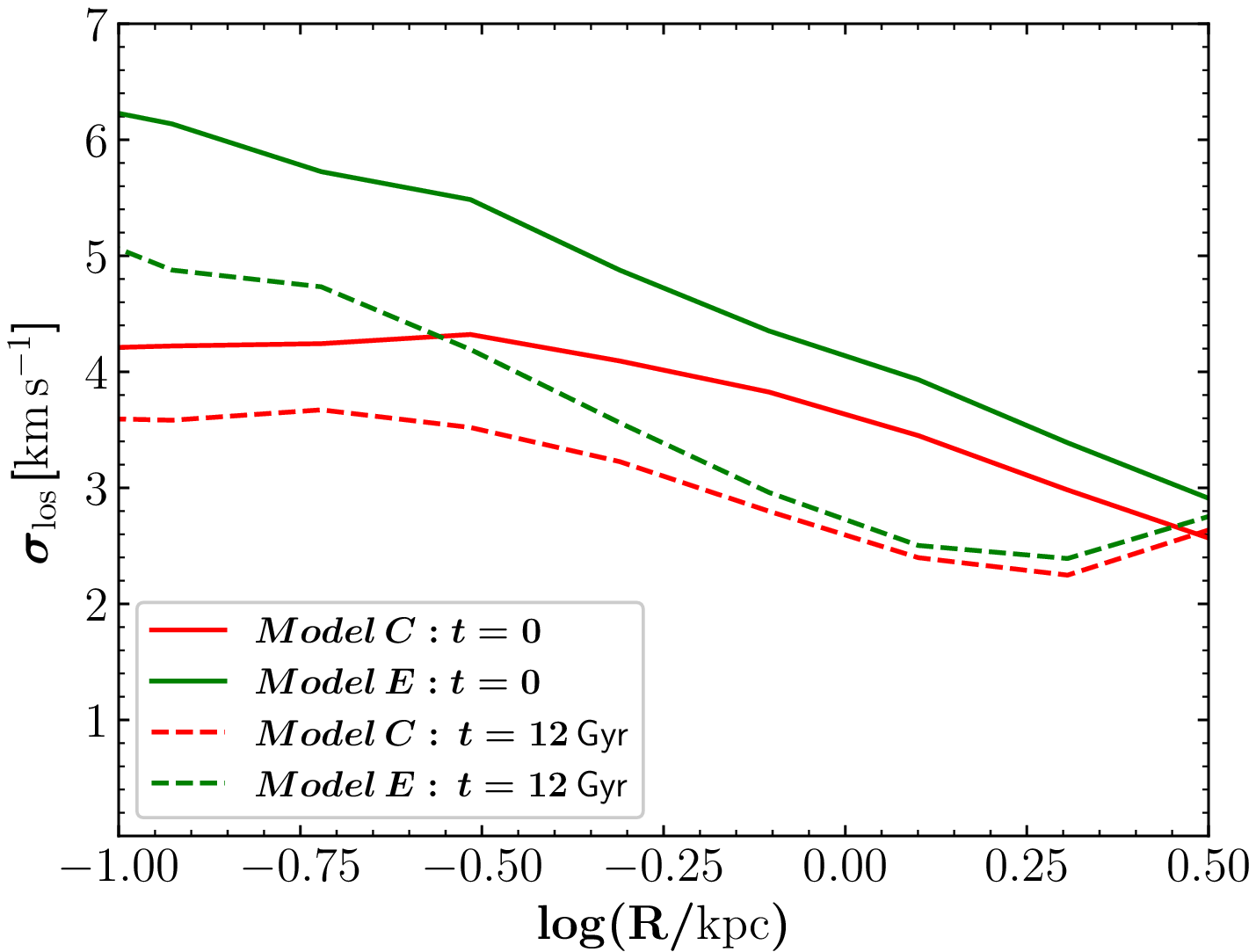,width=\hsize}}
  \caption{{Line-of-sight (along the $y$ axis) stellar velocity
    dispersion profile for the satellite at the beginning and
    at the end of the simulation for models C and E.}}
\label{fig:sigmalossat}
\end{figure}

It is also interesting to assess how the kinematics of the stellar
tidal tails depends on the initial stellar density distribution. For
this purpose we consider the $t=12\Gyr$ snapshot and, taking again the
$y$ axis as line of sight, we distinguish the leading tail (lying
above and to the right of the satellite in the bottom panels of
\Fig\ref{fig:snap}) and the trailing tail (lying to the left of the
satellite in the bottom panels of \Fig\ref{fig:snap}).  Specifically,
we assign to the leading tail all particles with $R>3\kpc$ and
$z>z_0-1.8(x-x_0)$, and to the trailing tail all particles with
$R>3\kpc$ and $z<z_0 -1.8(x-x_0)$, with $x_0=31.9\kpc$ and
$z_0=-124.8\kpc$.
$\sigmalos$ as a function of $R$ is shown in
\Fig\ref{fig:sigmalostail} for the leading and trailing tails of
models C and E. For given model, the two tails have similar
$\sigmalos$ profiles out to $R\approx 40\kpc$: at larger distances
form the satellite the leading tail tends to have higher stellar
velocity dispersion than the trailing tail.  For given tail (leading
or trailing), the $\sigmalos$ profile is systematically higher for
model E than for model C, which reflects the higher velocity
dispersion of the stellar component of model E in the initial
conditions (\Fig\ref{fig:sigmalossat}). To quantify the overall
velocity dispersion of each tail, we compute the quantity
$\sigmalostail$, defined by
\begin{equation}
  \sigmalostail^2=
  \frac{\sum_{j=1}^{\Nbin}\sigmalosjsquared\Sigmastarj}{\sum_{j=1}^{\Nbin}\Sigmastarj},
\end{equation}  
where $\sigmalosj$ and $\Sigmastarj$ are, respectively, the
line-of-sight stellar velocity dispersion and stellar surface density
of the $j$-th radial bin of the tail (we used $\Nbin=24$ bins
uniformly spaced in $R$ between $R\simeq3\kpc$ and $R\simeq100\kpc$).
The leading tail has $\sigmalostail\simeq 2.8\kms$ for model C and
$\sigmalostail\simeq 3.6\kms$ for model E; the trailing tail has
$\sigmalostail\simeq 2.8\kms$ for model C and $\sigmalostail\simeq
3.5\kms$ for model E.

\begin{figure}
  \centerline{\psfig{file=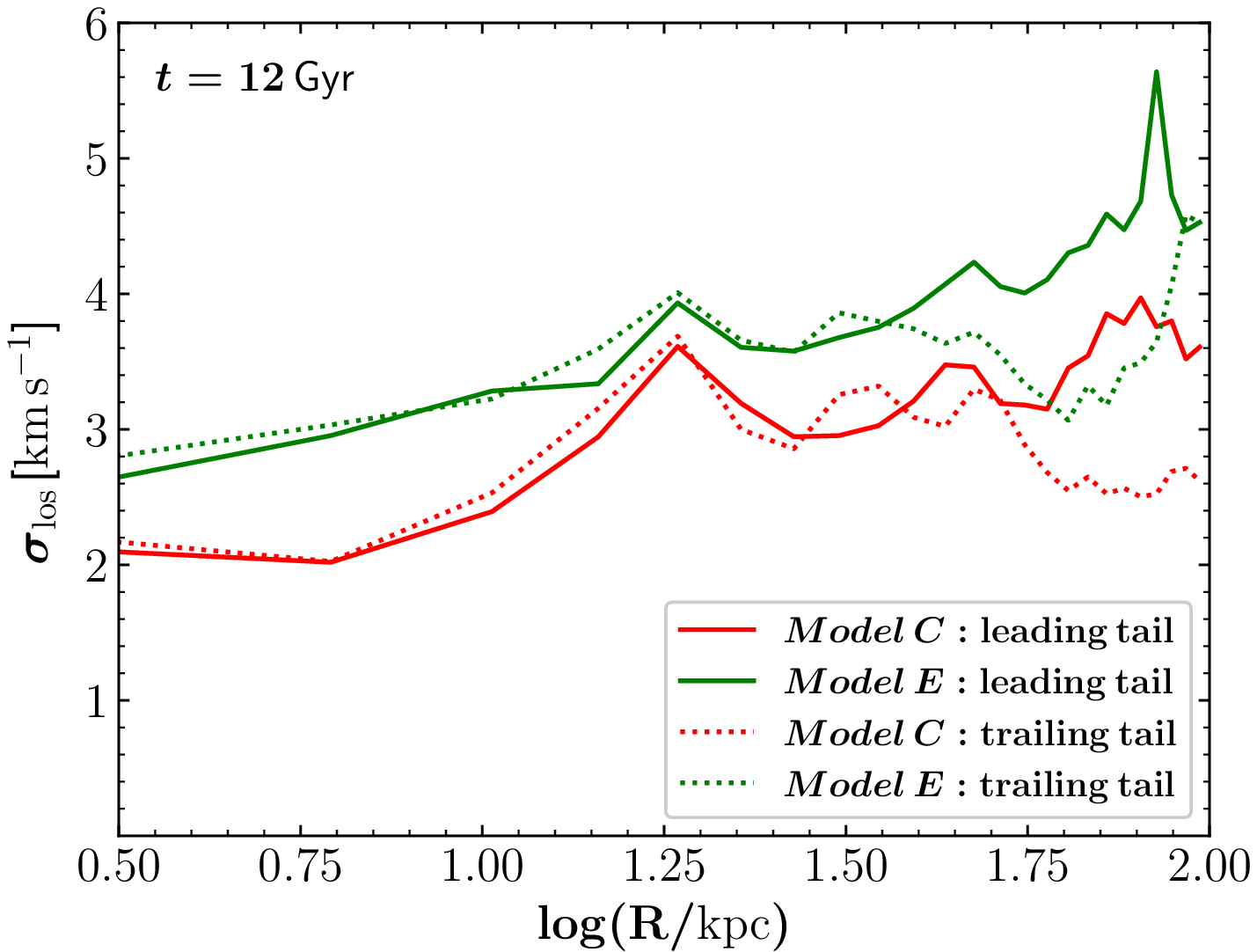,width=\hsize}}
  \caption{Line-of-sight (along the $y$ axis) stellar velocity
    dispersion profile of the leading and trailing tidal tails at the
    end of the simulation for models C and E.}
\label{fig:sigmalostail}
\end{figure}

\subsubsection{A family of models with smoothly varying $\Pstar$}
\label{sec:deponpstar}

So far we have applied to our simulation two models (C and E), that is
two choices of $\Pstar$. However, the power of the presented method
lies in the fact that infinite models can be explored by varying
continuously the values of the parameters of $\Pstar$. Thus we
illustrate here how some properties of the satellite and of the tails
vary in entire family of $n$ models whose extremes are models C and
E. The $i$-th member of this family of models (for $i=1,...,n$)
has $\Pstar(\E)$ given by \Eq(\ref{eq:por}) with parameters
\begin{equation}
\alpha=\alphaC+\frac{i-1}{n-1}(\alphaE-\alphaC),
\end{equation}
\begin{equation}
  \beta=\betaC+\frac{i-1}{n-1}(\betaE-\betaC),
\end{equation}
\begin{equation}
  A=\AC+\frac{i-1}{n-1}(\AE-\AC),
\end{equation}
and
\begin{equation}
  \Ezero=\EzeroC+\frac{i-1}{n-1}(\EzeroE-\EzeroC),
\end{equation}
where ($\alphaC$, $\betaC$, $\AC$, $\EzeroC$) and ($\alphaE$,
$\betaE$, $\AE$, $\EzeroE$) are the sets of values of parameters
of models C and E, respectively (see
\Sect\ref{sec:evostellardm}). With this definition we get model C for
$i=1$ and model E for $i=n$; for $1<i<n$ we get models with $\Pstar>0$
that are intermediate between models C and E: the stellar component is
more embedded in the DM halo for lower values of $i$. Each member of
this family of models can be conveniently labelled with the value of
its initial ($t=0$) stellar half mass radius $\rhalfstar$ (that is the
radius of the sphere containing half of the stellar mass), which
increases monotonically with $i$.  The initial stellar density profile
of the simulated satellite is shown in \Fig\ref{fig:rhoembed} for
models C and E, and for three representative intermediate models,
labelled with their values of $\rhalfstar$.

\Fig\ref{fig:rhalf} shows the dependence on $\rhalfstar$ of some
global properties of the stellar component of the simulated satellite
for the family of models defined above. The upper panel of
\Fig\ref{fig:rhalf} plots the fraction of stellar mass lost (defined
as the stellar mass in particles more distant than 3 kpc from the
satellite's centre) as a function of $\rhalfstar$ after $5\Gyr$ and
$12\Gyr$ of evolution. The fraction of stellar mass lost increases
smoothly from more embedded (smaller $\rhalfstar$) to less embedded
(larger $\rhalfstar$) models.  The lower panel of \Fig\ref{fig:rhalf}
plots the line-of-sight stellar velocity dispersion $\sigmalostail$
(see \Sect\ref{sec:stellarkin}) of the leading and trailing tails as a
function of $\rhalfstar$. $\sigmalostail$, which is similar for the
two tails for a given model, increases smoothly with $\rhalfstar$: the
less embedded the initial stellar component, the higher the velocity
dispersion of the stellar streams.

\begin{figure}
 \centerline{\psfig{file=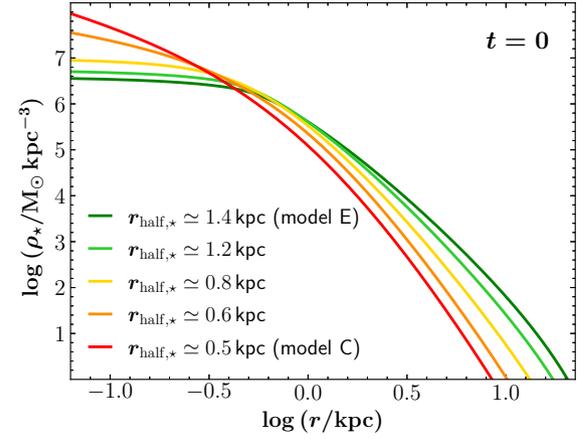,width=\hsize}}
  \caption{Initial stellar density profile of the simulated
    satellite for models C and E, and for three intermediate models,
    labelled with the value of their stellar half-mass radius.}
\label{fig:rhoembed}
\end{figure}

\begin{figure}
 \centerline{\psfig{file=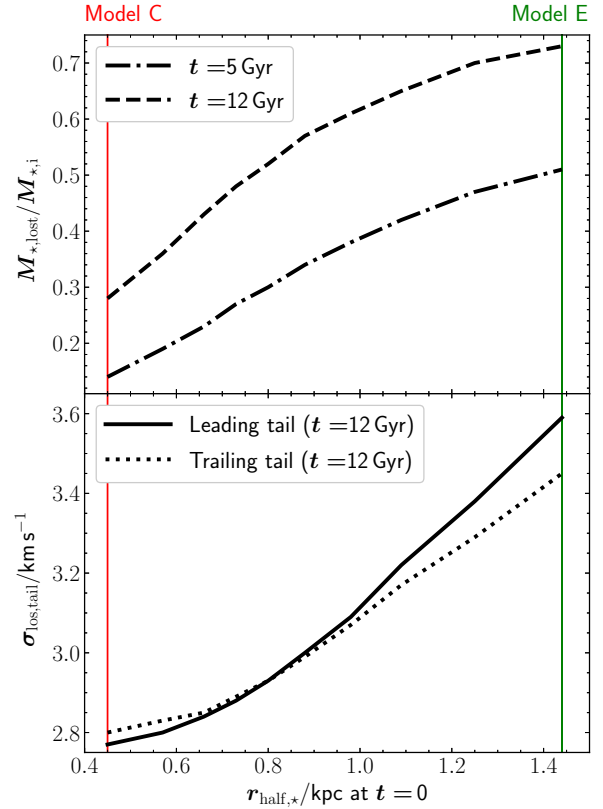,width=\hsize}}
  \caption{{\em Upper panel.} Fraction of stellar mass lost by the
    satellite in the simulation after $5$ and $12\Gyr$ of evolution as
    a function of the initial stellar half-mass radius for a family of
    models with smoothly varying initial stellar density distribution,
    ranging from the compact model C to the extended model E. Here
    $\Mstarlost$ is the stellar mass in particles more distant than
    $3\kpc$ from the satellite's centre and $\Mstari$ is the initial
    stellar mass. {\em Lower panel.} Final line-of-sight stellar
    velocity dispersion of the leading and trailing tails in the
    simulation for the same family of models as in the upper panel. }
\label{fig:rhalf}
\end{figure}

}

\section{Discussion and conclusions}
\label{sec:concl}

We have presented a new approach to $N$-body modelling of composite
collisionless stellar systems. The method, which we refer to as
effective multi-component $N$-body modelling, allows one to build a
one-component system, and interpret it {\em a posteriori} in infinite
ways as a multi-component system using functions of the integrals of
motion, dubbed portion functions. In an $N$-body simulation the
construction of the different components can be done in post
processing, thus greatly extending the applicability of the
simulation. As an example of application, we presented the results of
an $N$-body simulation of a satellite orbiting in the tidal field of
the Milky Way, which is interpreted {\em a posteriori} as a
two-component (stars plus DM) system. {This example nicely
  illustrates the potential of the presented method, by showing the
  dependence of the structure and kinematics of the final satellite
  and stellar streams on the choice of the portion function.}

For simplicity, we have presented as an application only the case in
which the parent one-component stellar system is spherical and
isotropic, and the portion function depends only on the initial
particle energy. But the very same method can be applied to
anisotropic spherical system as well as to non spherically symmetric
systems, provided their DF is known analytically or numerically. For
instance, one could build anisotropic multi-component spherical
systems with total DF $\ftot=\ftot(\E,L)$, where $L$ is the magnitude
of the angular momentum \citep[see][]{Bin08}, by using portion
functions $\Pk(\E,L)$. {A straightforward case is that of
  Osipkov-Merritt anisotropic spherical models \citep{Osi79,Mer85}, in
  which the DF is a function of a single variable $Q$, which is a
  combination of $\E$ and $L$, so $\Pk=\Pk(Q)$.} Moreover, the method
is not limited to spherical systems, and can be also applied to
axisymmetric systems with total DF $\ftot=\ftot(\E,\Lz)$, where $\Lz$
is the component of the angular momentum along the symmetry axis
\citep[see][]{Bin08}, using portion functions $\Pk(\E,\Lz)$, as well
as to both spherical and flattened models with total distribution
function $\ftot(\Jv)$ depending on the action integrals $\Jv$
\citep[e.g.][]{Bin14,Vas19}, and portion functions $\Pk(\Jv)$.

Of course, the presented effective $N$-body modelling method has its
own limitations. A necessary condition to use the effective modelling,
and thus to obtain the components' DFs by subtraction from the total
DF, is to know, numerically or analytically, the total DF, which can
be straightforward only in systems in which the total distribution is
simple, for instance because one of the components (typically the DM
halo) is dominant. Moreover, the construction of the portion functions
is relatively easy when the shapes of the system's components are
simple and similar among each other, but can be unfeasible in very
complex configurations. However, as it is well known, the build-up of
a complex composite stellar system (for instance an equilibrium galaxy
model with disc, bulge and non-dominant dark halo) is a hard task also
in standard approaches based on the DFs of the system's components.

The main power of the effective $N$-body modelling is that the
components of a composite simulated stellar system can be assigned in
post-processing.  This is especially useful when a simulation aims to
reproduce an observed distribution of stars, as it is often the
case. A typical case is that in which the composite system consists of
a stellar component and a DM halo.  For a given simulation, one can
{\em a posteriori} explore the space of the free parameters of the
stellar portion function (for instance the four-parameter space
$\alpha$, $\beta$, $A$ and $\Ezero$, when $\Pstar$ is in the form of
\Eq\ref{eq:por}) to find the set of parameters (and thus the initial
stellar and DM distributions) such that the final stellar distribution
represents best the observed data. In the near future we are going to
apply this approach to try to reproduce with $N$-body simulations the
observed properties of satellite dwarf spheroidal galaxies and
reconstruct their dynamical evolution and stellar mass loss history.

\section*{Acknowledgements}

FC acknowledges support from grant PRIN MIUR 20173ML3WW00 and from the INAF main-stream (1.05.01.86.31).

\section*{Data Availability}

The data underlying this article will be shared on reasonable request
to the corresponding author.

\bibliography{biblio_pstar}
\bibliographystyle{mnras}

\appendix

\end{document}